\documentclass[sigconf]{acmart}
\usepackage[utf8]{inputenc}
\usepackage{graphicx}
\usepackage[labelformat=simple]{subcaption}
\usepackage{enumitem}
\usepackage{array}
\usepackage{makecell}
\usepackage{booktabs}
\usepackage{multirow}
\usepackage{multicol}
\usepackage{tabularx}
\usepackage{arydshln}
\newcommand*\dhline{\specialrule{0pt}{1pt}{0pt}\hdashline[.4pt/3pt]\specialrule{0pt}{0pt}{2pt}}
\newcommand{\dcline}[1]{\specialrule{0pt}{1pt}{0pt}\cdashline{#1}[.4pt/3pt]\specialrule{0pt}{0pt}{2pt}}
\usepackage[disable]{todonotes} %
\usepackage{balance}
\usepackage{colortbl}
\usepackage{soul}

\usepackage[absolute,showboxes]{textpos}

\usepackage[skip=5pt]{caption}

\usepackage{tikz}
\usepackage{pgfplots}
\usetikzlibrary{trees,arrows,arrows.meta,positioning,fit,fadings,decorations,decorations.pathreplacing,decorations.text,backgrounds,calc,fit,math,quotes,external}
\usepgfplotslibrary{colorbrewer}

\usepackage{balance}
\usepackage{hyperref}
\usepackage{sistyle}

\copyrightyear{2022}
\acmYear{2022}
\setcopyright{acmlicensed}\acmConference[CoNEXT '22]{The 18th International Conference on emerging Networking EXperiments and Technologies}{December 6--9, 2022}{Roma, Italy}
\acmBooktitle{The 18th International Conference on emerging Networking EXperiments and Technologies (CoNEXT '22), December 6--9, 2022, Roma, Italy}
\acmPrice{15.00}
\acmDOI{10.1145/3555050.3569123}
\acmISBN{978-1-4503-9508-3/22/12}


\newcommand{\result}[1]{}

\definecolor{myred}{cmyk}{0, 0.7808, 0.4429, 0.1412}

\newcommand{\done}[1]{}

\definecolor{ColorGreen}{rgb}{0.59,0.73,0.38}
\definecolor{ColorRed}{rgb}{0.97,0.22,0.10}
\definecolor{ColorYellow}{rgb}{0.99, 0.88, 0.56}
\definecolor{ColorOrange}{rgb}{0.99, 0.50, 0.37}
\definecolor{ColorHSRed}{rgb}{0.655, 0.0, 0.0}
\definecolor{ColorHSOrange}{rgb}{1.0, 0.686, 0.231}
\definecolor{ColorHSLime}{rgb}{0.565, 0.933, 0.565}
\definecolor{ColorHSGreen}{rgb}{0.0, 0.502, 0.0}

\usepackage{pifont}

\usepackage{xspace}

\newcommand{\eg}{\textit{e.g.,}\xspace}
\newcommand{\ie}{\textit{i.e.,}\xspace}
\newcommand{\etc}{\textit{etc.}\xspace}
\newcommand{\one}{({\em i})\xspace}
\newcommand{\two}{({\em ii})\xspace}
\newcommand{\three}{({\em iii})\xspace}

\makeatletter
\renewcommand{\paragraph}[1]{\vspace*{0.03in}\noindent{\bf #1.}\hspace{0.25ex \@plus1ex \@minus.2ex}}
\newcommand{\paragraphNoDot}[1]{\vspace*{0.03in}\noindent{\bf #1}\hspace{0.25ex \@plus1ex \@minus.2ex}}
\makeatother

\newcommand*\circled[1]{\tikz[baseline=(char.base)]{
            \node[shape=circle,draw,inner sep=0.9pt] (char) {\sf \small #1};}}

\begin{document}

\setlength{\TPHorizModule}{\paperwidth}
\setlength{\TPVertModule}{\paperheight}
\TPMargin{5pt}
\begin{textblock}{0.8}(0.1,0.02)
	\noindent
	\footnotesize
	\centering
	If you cite this paper, please use the CoNEXT reference:
	M. Nawrocki, P. F. Tehrani, R. Hiesgen, J. Mücke,  T. C. Schmidt, and M. Wählisch.
	2022. On the Interplay between TLS Certificates and QUIC Performance.
	\emph{In Proceedings of CoNEXT ’22.}
	 ACM, New York, NY, USA, 10 pages. https://doi.org/10.1145/3555050.3569123
\end{textblock}

\date{}

\title[On the Interplay between TLS Certificates and QUIC Performance]{On the Interplay between TLS Certificates \\ and QUIC Performance}

\begin{abstract}

In this paper, we revisit the performance of the QUIC connection setup  and relate the design choices for fast and secure connections to common Web deployments.
We analyze over 1M~Web domains with 272k~QUIC-enabled services and find two worrying results.
First, current practices of creating, providing, and fetching Web certificates undermine reduced round trip times during the connection setup since sizes of 35\% of server certificates exceed the amplification limit.
Second, non-standard server implementations lead to larger amplification factors than QUIC permits, which increase even further in IP~spoofing scenarios.
We present guidance for all involved stakeholders to improve the situation.

\end{abstract}

\author{Marcin Nawrocki}
\email{marcin.nawrocki@fu-berlin.de}
\affiliation{%
  \institution{Freie Universit\"at Berlin}
  \country{Germany}
}

\author{Pouyan Fotouhi Tehrani}
\email{pft@acm.org}
\affiliation{%
  \institution{Weizenbaum Inst., Fraunhofer FOKUS}
  \country{Germany}
}

\author{Raphael Hiesgen}
\email{raphael.hiesgen@haw-hamburg.de}
\affiliation{%
  \institution{HAW Hamburg}
  \country{Germany}  
}

\author{Jonas M\"ucke}
\email{jonas.muecke@fu-berlin.de}
\affiliation{%
  \institution{Freie Universit\"at Berlin}
  \country{Germany}
}

\author{Thomas C. Schmidt}
\email{t.schmidt@haw-hamburg.de}
\affiliation{%
  \institution{HAW Hamburg}
  \country{Germany}  
}

\author{Matthias W\"ahlisch}
\email{m.waehlisch@fu-berlin.de}
\affiliation{%
  \institution{Freie Universit{\"a}t Berlin}
  \country{Germany}
}

\renewcommand{\shortauthors}{Nawrocki, et al.}

\maketitle

\section{Introduction}
\label{sec:introduction}

\begin{figure}
	\scriptsize
	\begin{tikzpicture}[
	>=Latex,
	box lab/.style={fill=white,inner ysep=-2pt,anchor=west,xshift=2pt,font={\bfseries}},
	clientserver/.style={fill=gray!40,minimum width=1cm, minimum height=2.25cm},
	msg lab/.style={inner ysep=0}
	]

	\newlength\iconh
	\pgfmathsetlength{\iconh}{30px}
	\newlength\arrowdist
	\pgfmathsetlength{\arrowdist}{9px}
	\newcommand{\boxbg}{gray!10}

	\node[clientserver, align=center] (client) {Client\\ (QUIC)};
	\node[clientserver,right=\columnwidth of client.west,anchor=east,align=center] (server) {Server \\ (QUIC)};

	\draw[thick,->] ($(client.north east)+(\arrowdist,-5pt)$) coordinate (init start) -- node[msg lab,fill=\boxbg] {Initial Message (\texttt{Client Hello})} ($(init start -| server.north west)+(-\arrowdist,0)$) coordinate (init end);
	\draw[thick,->] ($(init end)+(0,-\arrowdist)$) coordinate (init2 start) -- node[msg lab,fill=white] {Initial Message (\texttt{ACK, Server Hello})} ($(init start)+(0,-\arrowdist)$) coordinate (init2 end);
	\draw[thick,->] ($(init2 start)+(0,-0.9\arrowdist)$) coordinate (handshake start) -- node[msg lab,fill=white] {Handshake Message (TLS)} ($(init2 end)+(0,-0.9\arrowdist)$) coordinate (handshake end);

	\draw[thick,->] ($(handshake end)+(0,-1.1\arrowdist)$) coordinate (init ack start) -- node[msg lab,fill=\boxbg] {Initial Message (\texttt{ACK})} ($(handshake start)+(0,-1.1\arrowdist)$) coordinate (init ack end);
	\draw[thick,->] ($(init ack start)+(0,-0.8\arrowdist)$) coordinate (handshake ack start) -- node[msg lab,fill=\boxbg] {Handshake Message (\texttt{ACK, TLS})} ($(init ack end)+(0,-0.8\arrowdist)$) coordinate (handshake ack end);

	\draw[thick,->] ($(init ack end)+(0,-2.8\arrowdist)$) coordinate (handshake tls start) -- node[msg lab,fill=\boxbg] {Handshake Message (TLS)} ($(init ack start)+(0,-2.8\arrowdist)$) coordinate (handshake tls end);

	\begin{scope}[on background layer]
		\node[draw,fill=\boxbg,rounded corners,densely dotted,label={[box lab]north west:1-RTT},fit=(init start) (init ack end) (handshake ack end),inner ysep=5pt,inner xsep=7pt] (rt) {};

		\node[draw,fill=\boxbg,rounded corners,densely dotted,label={[box lab]north west:{Multi-RTT (if server data $> 3 \times$ \texttt{Client Initial})}},fit=(handshake tls start) (handshake tls end),inner ysep=6pt,inner xsep=7pt] (rt) {};

		\node[draw,ultra thin,rounded corners,fill=white,fit=(init2 end) (handshake start),inner ysep=6pt] (server msgs) {};
	\end{scope}

	\node[below=55pt of server msgs.east,anchor=east,xshift=-6] (3xmsg) {\shortstack[r]{\bf Should be $\leq 3 \times$ \texttt{Client Hello}.\\\bf Mainly steered by TLS cert chain of server.}};
	\draw[->] (server msgs.east) to[out=-20,in=0,distance=12] (3xmsg.east);
\end{tikzpicture}
	\caption{In QUIC handshakes, server replies are limited to 3$\times$ the size of the client \texttt{Initial} until the client is verified.}
	\label{fig:quictimesthree}
	\vspace{-0.5cm}
\end{figure}
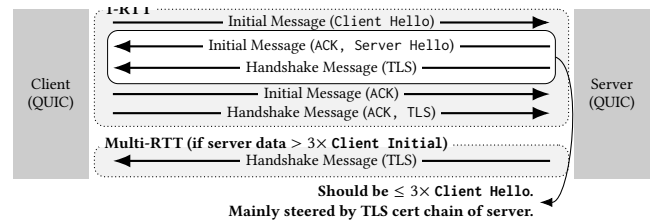

The QUIC protocol~\cite{RFC-9000} was designed to improve Web performance and reduce access latency~\cite{cllwk-itqda-17, sppb-eqpwc-21} while keeping communication confidential~\cite{vbkt-tbses-18}. 
A key approach is the reduction of initial roundtrip times by integrating the QUIC handshake with the TLS 1.3 handshake and coalescing multiple QUIC packets into one UDP~datagram. At the same time, security concerns about the UDP-based QUIC protocol demanded to limit the amplification potential, \ie the byte ratio of the server answer to the client request, for the initial reply to an (unauthenticated) client. 

In QUIC, the details of a connection handshake depend on a variety of factors: version negotiation, QUIC retry option, packet coalescing, and the size and compression of TLS certificates.
In this work, we focus on the latter because it has great relevance for the handshake process, see \autoref{fig:quictimesthree}. 
Using active and passive measurements we observe significant effects on latency, amplification, and protocol behavior in current deployments.

In detail, we contribute the following.
\begin{enumerate}
  \item Background on the interplay between the QUIC handshake and TLS certificates and prior work (\autoref{sec:background}).
  \item A measurement method to systematically analyze the problem space (\autoref{sec:method}).
  \item Analysis of QUIC server behaviors for different sizes of client \texttt{Initial} messages.
  The majority of QUIC~servers incorrectly amplify handshakes or require multiple RTTs, even for common \texttt{Initial} sizes used by Web browsers~(\autoref{sec:analysis:classifying-handshakes}).
  \item An in-depth study of QUIC handshake behavior that shows that multi-RTT handshakes are caused by large certificates and missing packet coalescence.
  Furthermore, some certificates unnecessarily contain cross-signed certificates instead of self-signed versions or include their trust anchors~(\autoref{sec:analysis:impact-certs}). 
  \item Empirical results highlighting the benefits of certificate compression during the handshake.
  99\% of certificate chains would remain below the allowed amplification factor (\autoref{sec:analysis:impact-certs}).
  \item A major reason why large amplification factors may appear during connection setups in the wild.
  For large CDNs, we observe up to $45\times$ amplification for spoofed handshakes~(\autoref{sec:analysis:amplification-potential}).
  \item Guidance to improve the situation (\autoref{sec:discussion}) and our artifacts, which are publicly available (\autoref{apx:artifacts}).
\end{enumerate}

\section{Background \& Related Work}
\label{sec:background}
In this section, we briefly recap the QUIC protocol mechanics of the connection setup, introduce challenges of TLS~certificates that undermine fast setups, and discuss related work.

\paragraph{The QUIC handshake and amplification mitigation}
QUIC~\cite{RFC-9000} was designed to provide low latency, reliability, and security on top of UDP.
A crucial part is the initial connection setup, which should be fast~\cite{wrwh-ppwop-19} and prevent attacks related to amplification~\cite{r-ahrnp-14,rowrs-adads-15} or state exhaustion~\cite{nawrocki2021quic}.
For this purpose, the QUIC handshake integrates TLS within the protocol handshake.
A client starts with an \texttt{Initial} that is answered by an \texttt{Initial} from the server and a \texttt{Handshake} packet that can be sent in a single UDP datagram (\emph{packet coalescence}).
The client confirms receipt with an \texttt{Initial ACK} and then sends an its \texttt{Handshake} message.
The server resends unconfirmed \texttt{Initial} and \texttt{Handshake} packets.
Protection against state exhaustion is achieved when a server uses \texttt{RETRY} packets but such protection is rarely deployed~\cite{nawrocki2021quic}.

To prevent amplification attacks, a server must not reply with more bytes than the QUIC \emph{anti-amplification factor} allows until the client IP~address is verified.
RFC 9000~\cite{RFC-9000} limits the data size from the server to 3$\times$ the bytes that have been received in the client \texttt{Initial}, see~\autoref{fig:quictimesthree}, and includes padding and resent bytes~\cite{RFC-9002}.
After the server validates the client by a complete roundtrip, it is free to send any amount of data.
The factor of three is low compared to the amplification potential of other protocols~\cite{r-ahrnp-14,rowrs-adads-15}.
We recap the IETF design of the threshold in more detail in~\autoref{apx:ietf-limit}.

\paragraph{QUIC TLS connection setup}
QUIC integrates TLS~1.3~\cite{RFC-8446} to cater for authenticated confidentiality and integrity~\cite{RFC-9001}.
A TLS 1.3 handshake is initiated by a client sending its supported cipher suites, key parameters, and other metadata in the first \texttt{Initial} message to the server, which in turn replies with its own parameters and an X.509 certificate~\cite{RFC-5280} used to authenticate its identity~\cite[\S4.4.]{RFC-9001}.

\begin{figure}
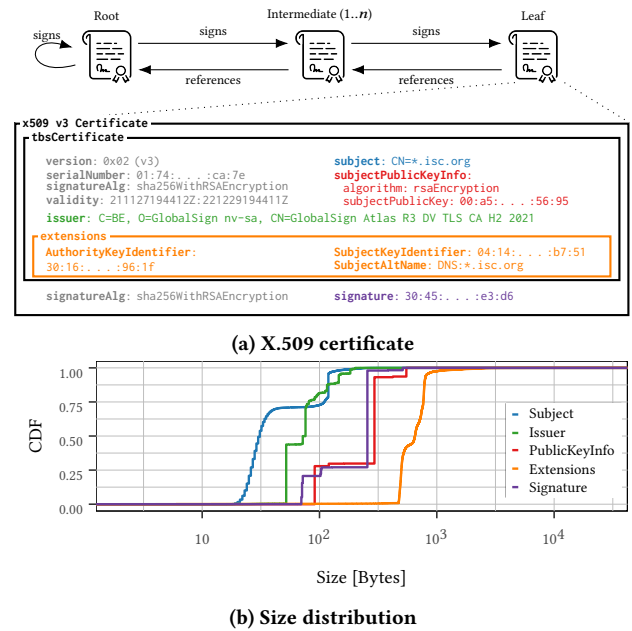

	\centering
	\tiny
	\begin{subfigure}{\columnwidth}
		\centering
		\input{./figures/tikz/x509-overview}
		\caption{X.509 certificate}
		\label{fig:cert-fields}
	\end{subfigure}
	\hspace{6pt}
	\begin{subfigure}{\columnwidth}
		\centering
		\input{./figures/tikz/field-size-cdf}
		\caption{Size distribution}
		\label{fig:cert-field-sizes}
	\end{subfigure}

	\caption{Example of a TLS certificate and our observed distribution for various X.509 certificate field sizes.}
	\label{fig:intro}
\end{figure}

In contrast to TLS over TCP, the sum of \emph{first responses} of a QUIC server must not be larger than the anti-amplification factor. %
This limit reduces the amplification attack surface but poses a challenge for benign QUIC peers to achieve the goal of low latency and low connection overhead.
Either the client sends an \texttt{Initial} that is large enough to allow the server to accommodate its reply within the anti-amplification limit, or the server adapts responses to be small enough.
Please note that the size of a server reply is mainly determined by its certificate, \ie the complete certificate chain sent by the server.
\autoref{fig:cert-fields} illustrates the structure of a TLS certificate.

Popular browsers use between 1250 and 1357~bytes in the \texttt{Initial} message, which can easily conflict with common sizes of server certificate (see \autoref{sec:analysis} for details).
Depending on public key and signature algorithms in use, sizes of issuer and subject names, as well as extensions (\eg subject alternative names), the total size of a certificate may vary by an order of magnitude.
\autoref{fig:cert-field-sizes} depicts the size distribution from our data corpus; certificate extension fields followed by signature and public key fields are the most space consuming in certificates.
A server can apply optimizations to its own certificate sizes, but it has no control over intermediate certificates that it delivers as part of the certificate chain of trust.
To compress the entire chain, TLS 1.3 provides certificate compression~\cite{RFC-8879}.
To take effect, client and server must support compression.
While the adoption of TLS 1.3 is well studied~\cite{Holz2020TLS}, analysis of certificate compression deployment is not included in prior research.

\paragraph{QUIC performance and adoption}
QUIC provides good performance~\cite{Biswal2016Faster, Cook2017Better, Carlucci2015HTTP, Kakhki2017Long} and can outperform TCP. %
Prior work suggests that some security trade-offs were specifically made in favor of improved latency~\cite{Lychev2015Secure}.
The handshake, however, can suffer from additional latency if client and server do not agree on a version directly~\cite{Gagliardi2020Sessions}.

Prior deployment studies mainly focus on the availability of QUIC~services.
QUIC adoption started before the finalization of the standard~\cite{mtmbb-aaqmd-20,rgdmm-fyewi-20,smp-wfaq-21} and continues since then~\cite{zirngibl2021quic}, led by hypergiants~\cite{rueth2018quic,mnhsz-wqomu-22}. %
DNS over QUIC lacks wide adoption and exhibits inefficient handshakes due to large certificates if \texttt{Session Resumption} is not used~\cite{kosek2022doq, kosek2022quicwebperf}.
Most closely related is \cite{Fastly2020Compression}, showing that 40\% of QUIC handshakes with uncompressed certificates may trigger an additional roundtrip, based on data from a specific CDN.
To the best of our knowledge, this paper is the first that systematically assesses the impact of TLS~certificates on QUIC~performance, leveraging comprehensive measurements.

\section{Measurement Method and Setup}
\label{sec:method}

We search for
\one common HTTPS services and
\two QUIC-services, to collect related TLS~certificates and compare performance of protocol design and deployment choices.
In this paper, we use the term \textit{service} to quantify the number of domains served via a specific protocol, irrespectively of whether these domains are delivered by the same IP host, \ie we present a domain-centric perspective.
The point of departure for our scans is the Tranco list~\cite{LePochat2019tranco} generated on September 10, 2022, since the Tranco list provides a good compromise~\cite{shgjj-lwtss-18} between reflecting  popularity and robustness.
Subsequently, we scan 1M domain names to broadly capture what clients receive when contacting a web domain.

When conducting our measurements, we leverage existing tools where possible and minimize extensions to achieve maintainability and ease reproducibility.
Unfortunately, there is no single~tool available that implements all necessary features.
We present an overview of our toolchain in \autoref{apx:artifacts}, \autoref{fig:toolchain}.

\subsection{TLS Certificate Scans via HTTPS}
Not all names in the Tranco list resolve to web servers that allow for TLS over TCP~connections.
For each name in the list, first, we try to resolve IPv4 addresses using Google public resolver \texttt{8.8.8.8}.
Upon success, we then try to establish HTTP connections on ports 80 and 443 and follow any redirects using HTTP(S) (status code 3xx) and HTML (\texttt{meta} tag with \texttt{http-equiv} attribute).
For every secure domain, including all redirects, we collect TLS~certificates.

We were able to resolve 976k (out of 1M).
For 13k names, the domain query returned \texttt{SERVFAIL}, 9k could not be resolved (\texttt{NXDOMAIN}), and the remaining either timed out ($10s$) or refused the answer (\texttt{REFUSED}~\cite{RFC-1035,RFC-6895}).
About 866k names returned an IP address (\texttt{A} record).
For every domain name which resolved to an IP address we tried to establish an HTTP connection on both ports 80 (HTTP) and 443 (HTTPS).
After following redirects, we collected 821k unique certificates for more than 1.1M names along the redirection path. %

\subsection{QUIC Scans}
We analyze QUIC handshakes in two scenarios: \one~a complete handshake including client verification and \two an incomplete handshake imitating an unverified client, \eg when clients spoof IP~addresses.

\paragraph{Complete handshakes}
We scan all domains discovered during our certificate collection via HTTPS and assign each successful handshake to one of the four groups:
\begin{enumerate}%
    \item \textcolor{ColorHSGreen}{\underline{\color{black}\bf 1-RTT} {\color{black}\bf (optimal):}} Handshakes that complete within 1-RTT and comply with the anti-amplification limit.
	\item \textcolor{ColorHSLime}{\underline{\color{black}\bf RETRY} {\color{black}\bf (less efficient):}} Handshakes that require multiple RTTs because the \texttt{Retry} option is used~\cite[\S8.1.]{RFC-9000}.
	\item \textcolor{ColorHSOrange}{\underline{\color{black}\bf Multi-RTT} {\color{black}\bf (unnecessary):}} Handshakes that do not use \texttt{Retry} but require multiple RTTs because of large certificates.
    \item \textcolor{ColorHSRed}{\underline{\color{black}\bf Amplification} {\color{black}\bf (not RFC-compliant):}} Handshakes that complete within 1-RTT but exceed the anti-amplification~limit.
\end{enumerate}

To conduct QUIC~handshakes and assign groups, we use \texttt{quic\-reach}~\cite{quicreach-github}, extended by \texttt{RETRY} support.
We find 272k QUIC services~($\sim$25\%).
To investigate the effect of client \texttt{Initial} sizes on server handshake behavior we vary the client \texttt{Initial} size between 1200 bytes (mandated minimum~\cite{RFC-9000}) and 1472 bytes (dictated by our MTU since QUIC forbids fragmentation) in steps of 10~bytes.
Handshakes targeting the same domain service pause 30~minutes to avoid side effects such as DDoS mitigation.

\texttt{quicreach} does not provide access to certificates nor does the underlying stack support certificate compression.
We rescan with
\one \texttt{QScan\-ner}~\cite{qscanner-github} to access TLS certificates sent over QUIC and
\two extend \texttt{quiche}~\cite{quiche-github} to support three popular TLS compression algorithms in QUIC.

In the majority of cases (96.7\%), we find that the same TLS certificates are used in both QUIC and HTTPS deployments for the same domains, which confirms prior work~\cite{zirngibl2021quic}.
For the remaining 3.3\% of QUIC services, certificates differ from TLS over TCP.
These differences are mainly due to certificate rotations during the period of time between our HTTPS and QUIC scans, leaving only 0.47\% of QUIC services with different certificates because of other reasons.
To sanitize inconsistent data, we decide to base our QUIC~certificate analysis on the TLS certificates gathered via HTTPS.

\paragraph{Incomplete handshakes}
To analyse the performance when a client successfully initiates but does not complete a handshake, \eg because of malicious activities, we conduct two measurements.
First, we collect QUIC backscatter from a telescope during January 2022.
Since telescopes do not emit any traffic, we can observe server behavior to non-responding, spoofed client IP~addresses.
Here, we group QUIC traffic by major content providers and source connection IDs (SCIDs).
Second, we send a single  \texttt{Initial} of 1252~bytes to the servers without sending \texttt{ACK} messages, using ZMap~\cite{dwh-zfiss-13}.

\section{Results}
\label{sec:analysis}

In this section, we
\one present our analysis of complete handshakes,
\two study TLS~certificates as potential reason for performance drawbacks in more detail, and
\three show results that reveal QUIC amplification potentials in the wild.

\subsection{Classifying QUIC Handshakes}
\label{sec:analysis:classifying-handshakes}

\begin{figure}[t]
	\begin{center}
		\includegraphics[width=1\columnwidth]{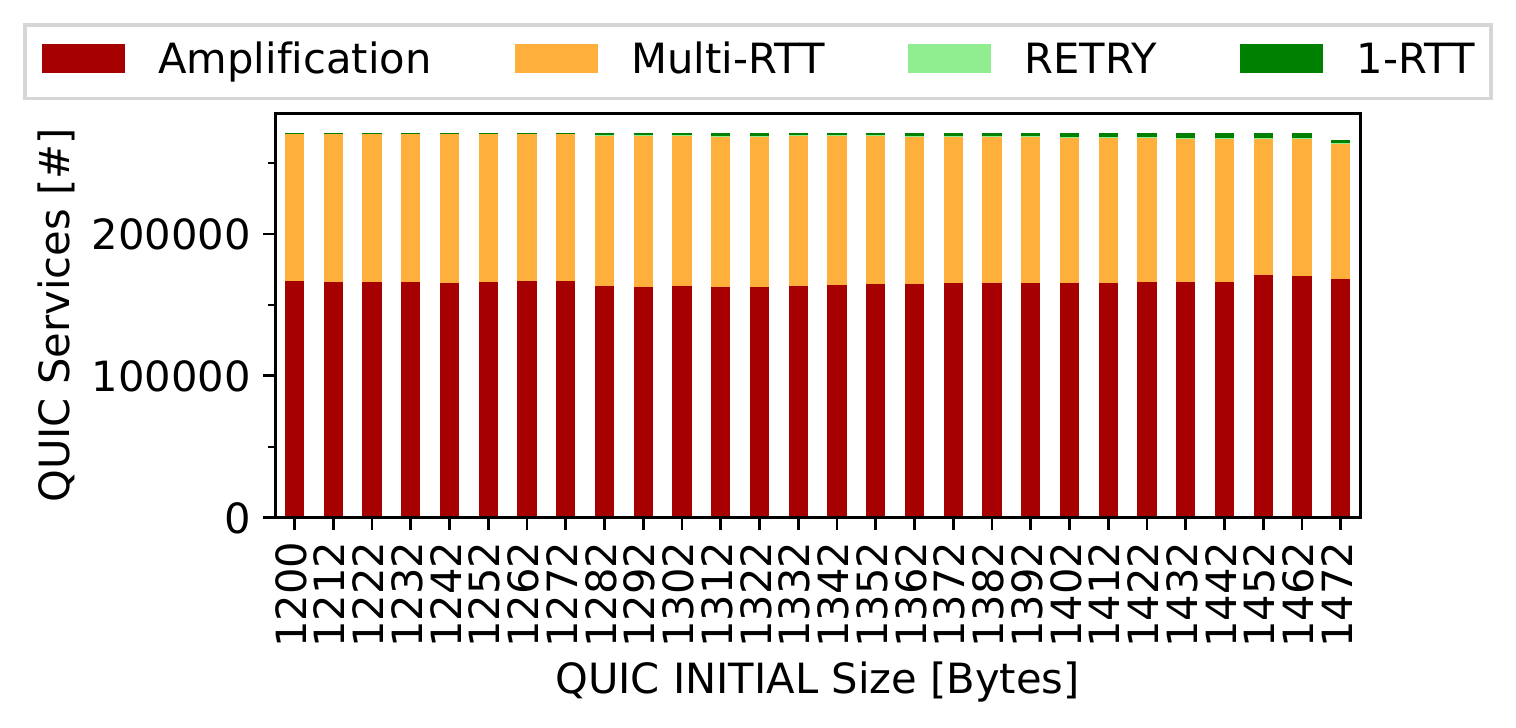}
		\caption{Influence of QUIC \texttt{Initial} sizes on the QUIC handshake. With respect to all names, we find almost no effect.}
		\label{fig:quic-handshake-classes-abs}
	\end{center}
\end{figure}

\begin{table}%
	\caption{Comparison of QUIC INITIAL packet sizes and support for TLS 1.3 certificate compression in popular browsers.}
	\label{tab:browersize}
	\footnotesize
	\centering
	\begin{tabular}{l >{\ttfamily}r r c r r}
		\toprule
		                   &                       & Init. Size & \multicolumn{3}{c}{Compression} \\
		\cmidrule{4-6}
		Browser            &  \textnormal{Version} & [Bytes]    & Algorithm$^3$   & Rate$^4$ & Services$^5$\\
		\midrule
		Firefox            &  101.x              & 1357       & --              & --       & --      \\ \dhline
		Chromium-based$^1$ &  105.x         & 1250$^2$   & brotli          & 73\%     & 96\%    \\ \dhline
		Safari (macOS)     &  15.5                 &  no QUIC   & \texttt{zlib}   & 74\%     & 0.05\%  \\ \dhline
		                   &                       &            & \texttt{zstd}   & 72\%     & 0.05\%  \\
		\bottomrule
	\end{tabular}
	\smallskip
	\begin{minipage}{8.8cm}
		$^1$ Chrome \texttt{102.x}, Brave \texttt{V1.39}, Vivaldi \texttt{5.3.x}, Edge \texttt{102.x}, Opera \texttt{88.0.x}.

		$^2$ Recently reduced from 1350~\cite{google-initial-change-github}.
		$^3$ Tested with TLS~1.3 in TCP.
		\\
		$^4$ Mean rate observed by our Quiche client.
		$^5$ Out of 272k QUIC services.

	\end{minipage}
\end{table}

\paragraph{Overview}
\autoref{fig:quic-handshake-classes-abs} shows the absolute number of handshakes types for all QUIC-reachable names, depending on the \texttt{Initial} size.
For an \texttt{Initial} size of 1362~bytes, which is similar to common browser default values (see \autoref{tab:browersize}), we find that 61\% of handshakes are classified as amplifying and 38\% as requiring multiple RTTs.
Worryingly, the \texttt{Retry} and 1-RTT handshakes account for only 0.07\% and 0.75\%, respectively.
This means that a priori DoS protection and fast handshakes are rare, unveiling that the QUIC design goals have not been met in the wild,~yet.

We now investigate the effect of different \texttt{Initial} sizes.
We find that amplifying handshakes occur independently of the \texttt{Initial} size.
However, we observe an interdependence between multi-RTT and 1-RTT handshakes.
With larger \texttt{Initials}, multi-RTT handshakes are less likely and 1-RTT handshakes more likely (de- and increase by $\sim$1\%).
This nicely illustrates the performance impact of the interplay between \texttt{Initial} sizes and deployed certificate sizes.

We also observe that the reachability of QUIC services is reduced by 1.2\% for large \texttt{Initials}, as indicated by the decreased height of the stacked bars.
Interestingly, this effect is more pronounced for top-ranked services (not shown).
The top 1k and top 10k domains are seeing a 25\% and 12\% drop in reachability, respectively.
We argue that this corresponds to load-balancer deployments that are more likely to be used for very popular names.
Load-balancers utilize packet tunneling to distribute the load across multiple, redundant server instances.
Packet encapsulation used during tunneling adds bytes due to additional headers, which then exceed the local MTU.
Our observations of reachability issues comply with prior measurements~\cite{langley2017quic}.
Other than that, we find little differences across ranks, compare \autoref{apx:influence-rank}.

\begin{figure}[t]
	\begin{center}
		\includegraphics[width=1\columnwidth]{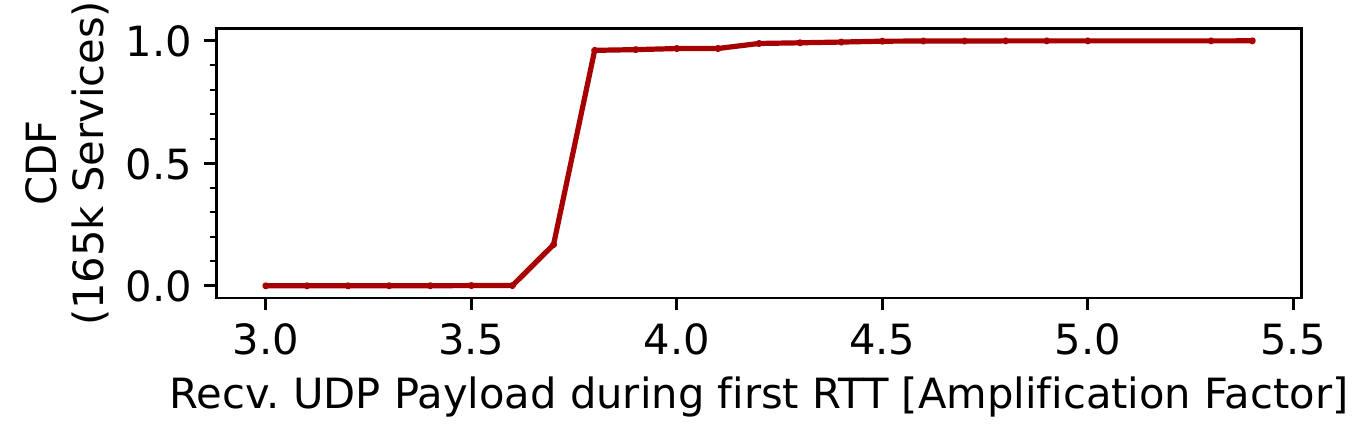}
		\caption{Amplification factor during first RTT. For complete client handshakes, the amplification is relatively small.}
		\label{fig:quic-ampl-factor-1rtt}
		\vspace{-0.5cm}
	\end{center}
\end{figure}

\paragraph{1-RTT exceeding anti-amplification limit}
Independently of the \texttt{Initial} size, the majority of handshakes exceed the anti-amplification limit in the first RTT.
We calculate the amplification factor for our default \texttt{INITIAL} scans of 1362~bytes by dividing UDP payload bytes received by the UDP payload bytes sent by the client.
\autoref{fig:quic-ampl-factor-1rtt} shows the amplification distribution.
The amplification factor, although exceeded, remains relatively small below \texttt{6x}.

\paragraph{Cloudflares missing coalescence explains amplification}
Based on TLS information and additional IP prefix mapping, we find that 96\% of the amplifying handshakes are completed with Cloudflare servers and subject to the same implementation behavior.
Surprisingly, we observe exactly $2462$ superfluous QUIC padding bytes for $\approx$157k handshakes.
In these cases, although the TLS data can vary in size, the remaining QUIC bytes are constant in size.
Cloudflare servers do not support packet coalescence at two levels:
\one \texttt{Initial} flags are sent separately, leading to two UDP datagrams. The first containing the \texttt{ACK} and the second the \texttt{ServerHello} flag, both of which are padded resulting in 2462 extra bytes, although only the latter elicits \texttt{ACKs} and thus requires padding.
\two We do not observe any coalescence of \texttt{Initial} and \texttt{Handshake} messages.
The extra bytes account for $\approx$60\% of the limit but are (incorrectly) not considered during amplification limit checks.
We report this implementation shortcoming to Cloudflare.

\paragraph{\texttt{Retry}}
We observe $\approx$200 services that predominantly request a \texttt{Retry} to authenticate client addresses.
We conclude that always-on DDoS mitigation is currently not widely adopted, however, \texttt{Retrys} might also be triggered adaptively based on the current server load.

\paragraph{Multi-RTT (no \texttt{Retry})}
Due to rare deployment of \emph{always-on} \texttt{Retry} messages, we assume that multi-RTT handshakes are caused by other factors.
We analyze these factors, \ie TLS certificates, in more detail in the next section.

\subsection{Impact of TLS Certificates}
\label{sec:analysis:impact-certs}

\begin{figure}%
	\begin{center}
		\includegraphics[width=1\columnwidth]{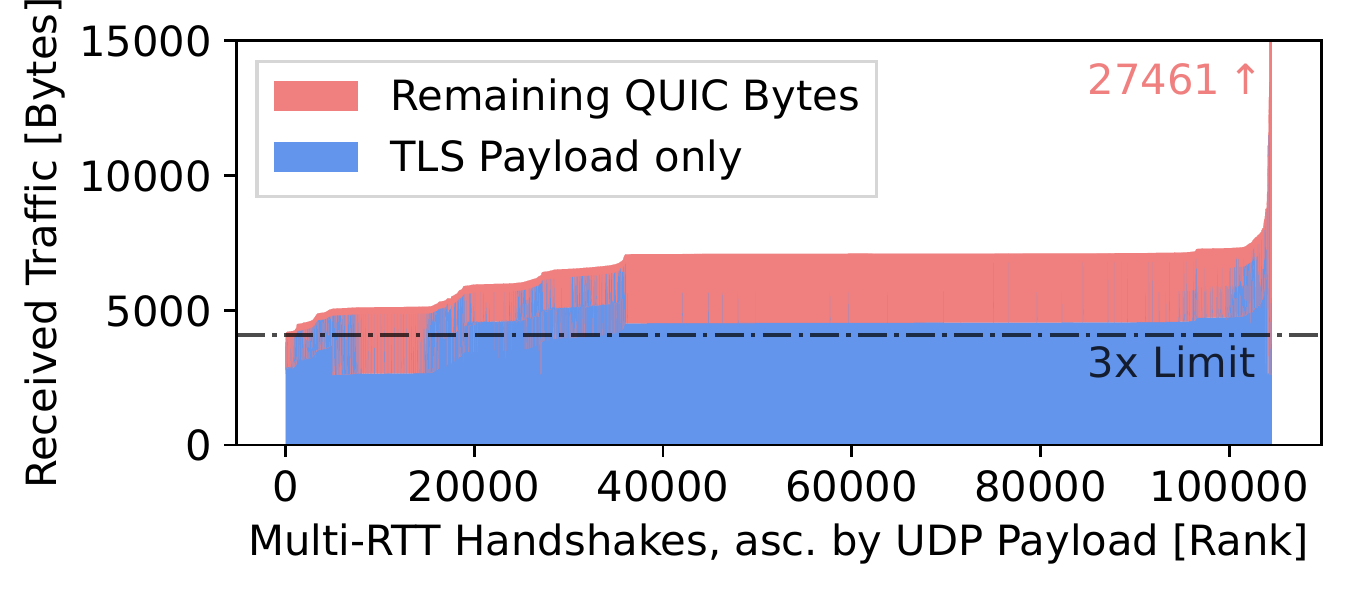}
		\caption{Payload exchanged during multi RTT handshakes. TLS bytes almost always exceed the limit but also QUIC padding can have a significant impact.}
		\label{fig:quic-tls-payload}
	\end{center}
\end{figure}

\begin{figure}%
	\begin{center}
		\includegraphics[width=1\columnwidth]{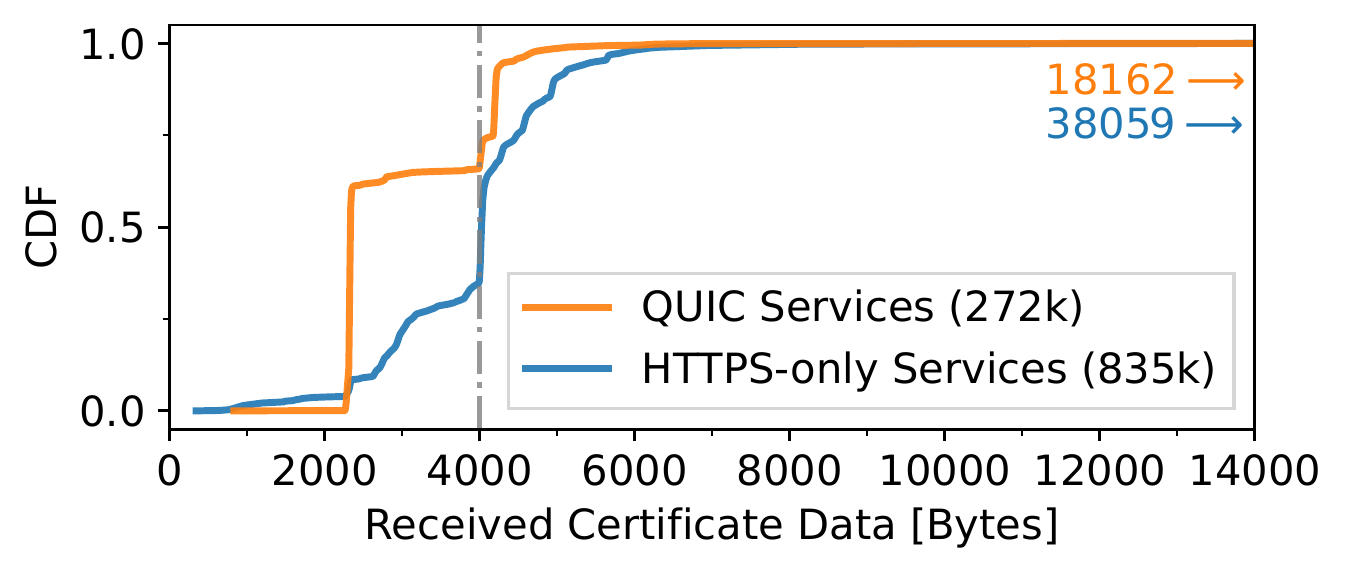}
		\caption{Distribution of certificate sizes grouped by QUIC support. QUIC domains use smaller certificates.}
		\label{fig:cert-cdf-size}
	\end{center}
\end{figure}

\begin{figure*}
	\begin{subfigure}[b]{.49\textwidth}
		\begin{center}
			\includegraphics[width=\textwidth,keepaspectratio]{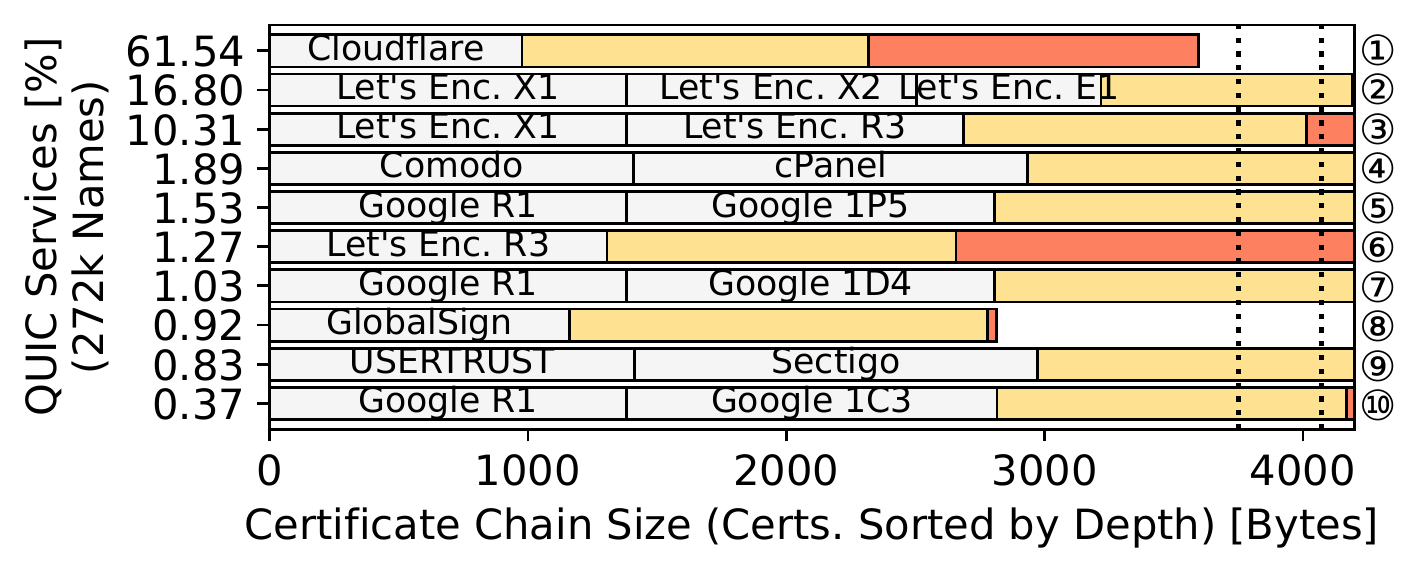}
			\caption{QUIC services}
			\label{fig:cert-chain-dependency-only-quic}
		\end{center}
	\end{subfigure}
	\begin{subfigure}[b]{.49\textwidth}
		\begin{center}
			\includegraphics[width=\textwidth,keepaspectratio]{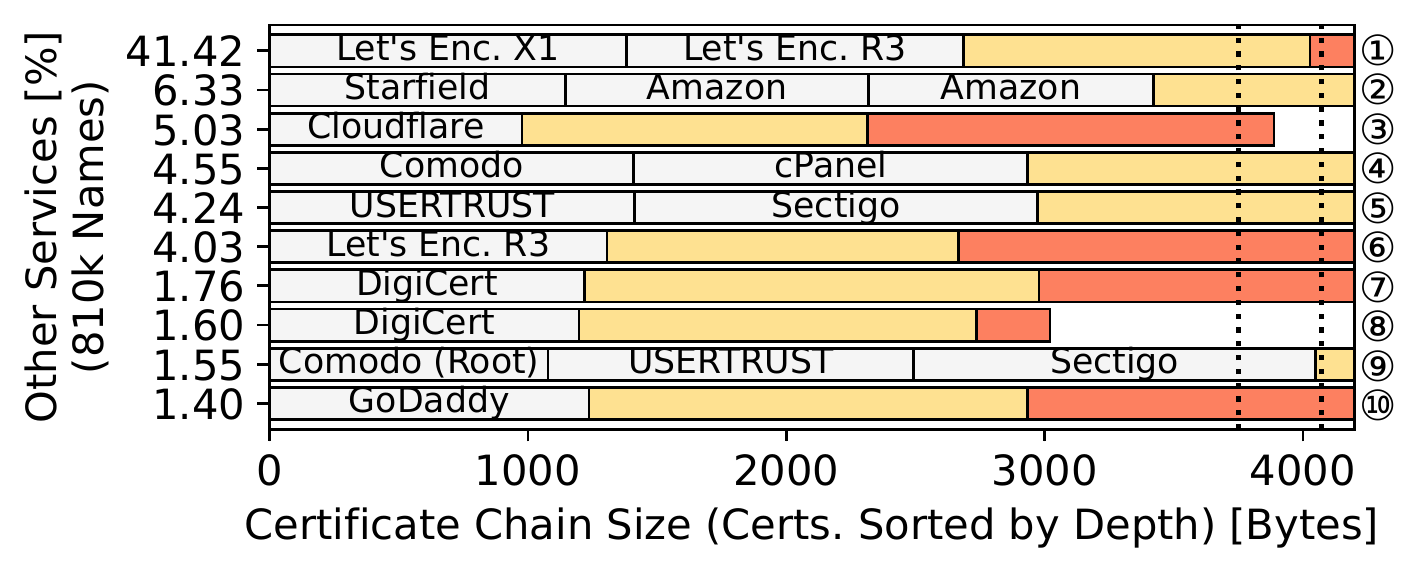}
			\caption{Only HTTPS services}
			\label{fig:cert-chain-dependency-no-quic}
		\end{center}
	\end{subfigure}
	\caption{Certificate chain sizes, depths and their dependency. {\color{ColorYellow}$\blacksquare$}~represents median leaf size, and {\color{ColorOrange}$\blacksquare$}~the additional bytes required for the maximum leaf size. Dotted lines represent the max allowed reply sizes of a server given common client \texttt{Initial} sizes. The x-axis is cut off at 4200 bytes. Average sized certificate chains are likely to exceed QUIC amplification limits.}
	\label{fig:cert-chain}
\end{figure*}

We presume that TLS certificate data causes multi-RTT handshakes.
To verify our assumption, we divide the bytes exchanged during a handshake into TLS payload and QUIC-related payload, \eg QUIC header and padding.

In the majority of cases (87\%), TLS payloads alone exceed the amplification limit (see \autoref{fig:quic-tls-payload}).
The distribution of (uncompressed) certificate chain sizes exchanged over TLS is shown in \autoref{fig:cert-cdf-size}.
Overall, we observe a median of 2329 bytes for certificate chains delivered by QUIC domains compared to 4022 bytes for other names.
We find that 35\% of all certificate chains exceed even the larger of the two common amplification limits (3$\cdot$1357 bytes).
This means that domains without QUIC support will be affected negatively when they decide to support QUIC in the future and continue to use existing certificates.

\begin{figure}%
	\begin{center}
		\includegraphics[width=1\columnwidth]{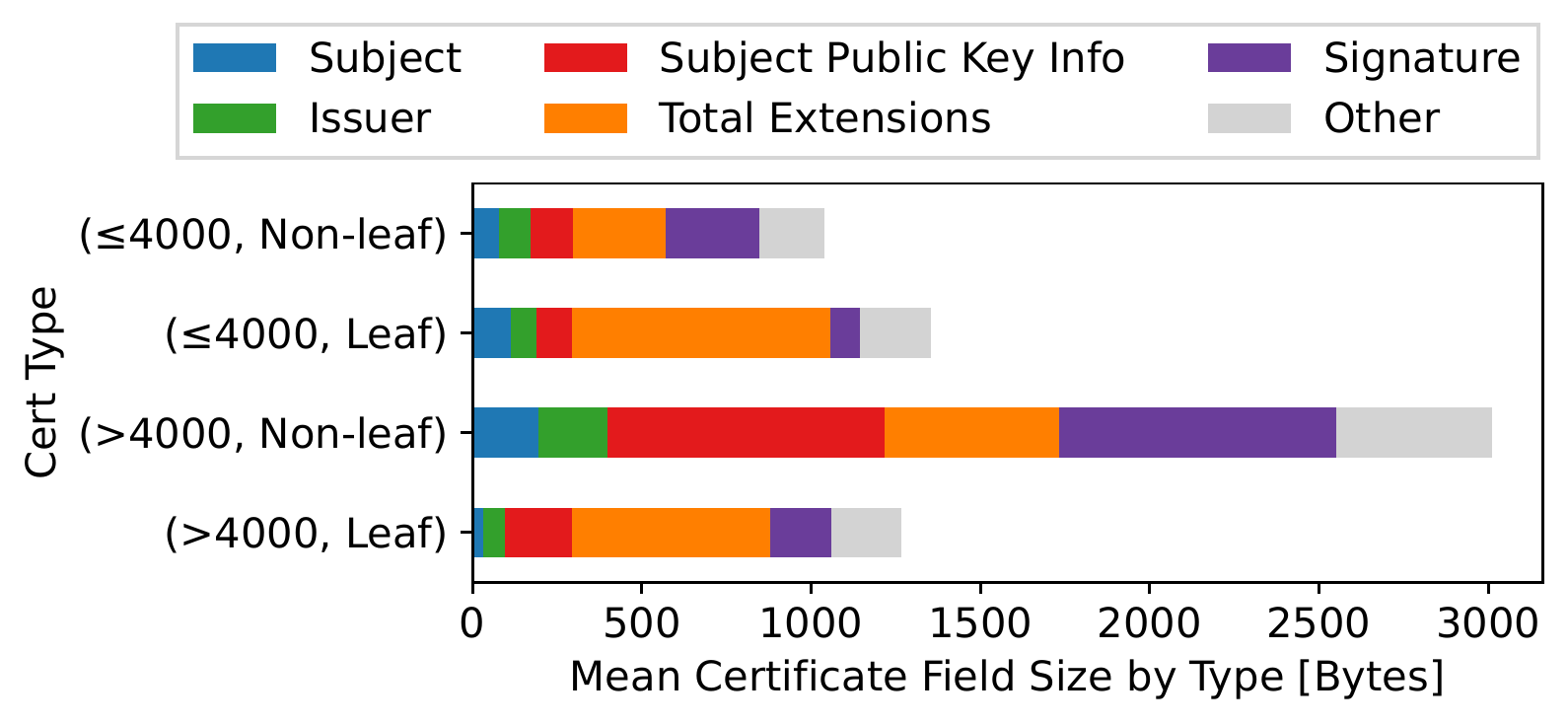}
		\caption{Mean sizes of certificate fields for QUIC domains. Non-leafs contribute most bytes to large chains.}
		\label{fig:cert-anatomy-size}
	\end{center}
\end{figure}

\paragraph{Popular parent certificates for QUIC unveil consolidation}
By zooming into certificate chains, we examine how the choice of a specific CA can impact the size of the certificate chain that a service provider needs to deploy.
For this analysis, we exclude certificate chains that are not ordered correctly.
\autoref{fig:cert-chain} exhibits the top-10 certificate chains deployed.
Each white box represents the sizes of the certificates in the chain (excluding leaf certificates), yellow boxes ({\color{ColorYellow}$\blacksquare$}) and orange boxes ({\color{ColorOrange}$\blacksquare$}) represent the median sizes and the largest leaf certificate that we observed in that chain.

Overall, we find that 7 out of 10 parent chains, together with the median leaf size, exceed common amplification limits (5 out of 10 for HTTPS-only services).

For both QUIC and non-QUIC services the shortest chains, \ie the smallest number of intermediates, are issued by Cloudflare followed by Let's~Encrypt~R3, GlobalSign, DigiCert, and GoDaddy.
We also observe cases in which cross-signed certificates are redundantly included in chains while the self-signed version of the same public key is already included in client trust stores.
For example, row~\circled{2} and \circled{3} in \autoref{fig:cert-chain-dependency-only-quic} include the cross signed version~\cite{crt-sh-3958242236, lets-encrypt-hierarchy} of \texttt{ISRG Root X1} (signed by \texttt{DST Root CA X3}) instead of relying on the self-signed variant~\cite{crt-sh-9314791, lets-encrypt-hierarchy},
as in row \circled{6}.
In other cases, servers superfluously include trust anchors (\ie root) certificates~(\eg row~\circled{9} in \autoref{fig:cert-chain-dependency-no-quic}).

Furthermore, a high consolidation trend for QUIC services is visible, as the top-10 parent chains cover 96.5\% of QUIC services.
For HTTPS-only services, this trend is less pronounced with only 72\% of services.
Consequently, to improve the deployment of QUIC~services, optimizing the parent chains can have a significant, beneficial effect but only needs to involve a small number of stakeholders.
Certificates delivered by QUIC~servers tend to use more efficient crypto algorithms, though, compared to non-QUIC Web services (see \autoref{tbl:crypt-dist}).

\paragraph{Non-leaf certificates bring the heavy load}
We find very large certificate chains requiring transmissions between 18k and 38k bytes, indicated by the long tail above 4000 bytes in \autoref{fig:cert-cdf-size}.
We proceed to use this value as a threshold to classify certificate chains.

\autoref{fig:cert-anatomy-size} depicts the mean size of various TLS certificate fields divided into leaf and non-leaf certificates. %
We observe that for large chains the sum of public key and signature sections on non-leaf certificates has the biggest impact on the chain size.
This again shows the negative effects of selecting a large non-leaf parent chain, even if the related leaf certificate has a reasonable size.

We also find that large cruise-liner leaf certificates~\cite{cangialosi2016httpskeys} are rarely used in QUIC deployments, details see \autoref{apx:cruise-liner}.

\begin{table}%
	\caption{Relative ratio of crypto algorithms and key lengths [bits] in use (limited to types with a frequency of~$>1\%$). HTTPS-only domains depend heavily on RSA.}
	\label{tbl:crypt-dist}
	\footnotesize
	\begin{tabularx}{\columnwidth}{l X r r r r}
		\toprule
		& & \multicolumn{2}{c}{RSA} & \multicolumn{2}{c}{ECDSA} \\
		\cmidrule(r){3-4} \cmidrule{5-6}
		Service & Certificate & 2048 & 4096 & 256 & 384 \\
		\midrule
		\multirow{2}{*}{QUIC} & Non-leaf & 15.1\% & 22.4\% & 40.4\% & 22.1\%\\
		\dcline{2-6}
		& Leaf & 19.2\% & 1.4\% & 78.9\% & 0.0\%\\
		\midrule
		\multirow{2}{*}{HTTPS-only} & Non-leaf & 63.3\% & 32.1\% & 2.7\% & 1.6\% \\
		\dcline{2-6}
		& Leaf & 81.4\% & 8.1\% & 7.8\% & 1.9\% \\
		\bottomrule
	\end{tabularx}
\end{table}

\paragraph{Compression helps}
Compressing certificate chains can avoid exceeding anti-amplification limits and thus improve the situation in the future.
Our synthetic experiment of compressing collected certificate chains shows a median compression rate of $\approx$65\%.
This keeps the size under the amplification limits for 99\% of TLS chains, which in turn prevents multi-RTT QUIC handshakes.

We find that 96\% of QUIC services currently support the brotli algorithm, which is used by Chromium derivatives.
The support of multiple algorithms, however, is very rare with only 0.05\% of QUIC services offering all three.
These services relate to Meta.

The mean compression ratio in the wild is 73\%, which is close to our synthetic experiments.
Here, 99\% of all compressed certificates fit below a common anti-amplification limit (3$\cdot$1357~bytes).

\subsection{Examining Amplification Potential}
\label{sec:analysis:amplification-potential}

\begin{figure}[t]
  \begin{center}
  \includegraphics[width=1\columnwidth]{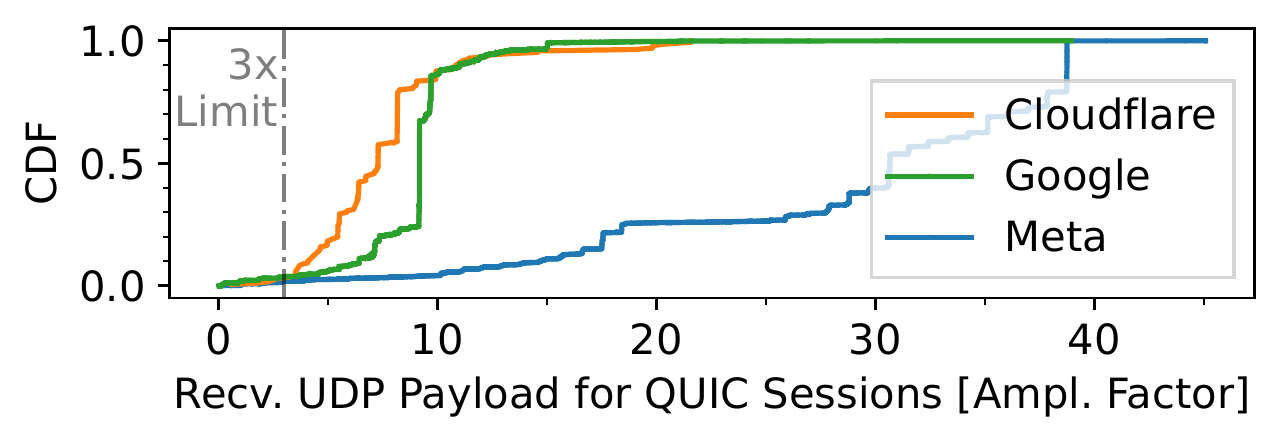}
      \caption{QUIC amplification factors including resends when clients do not respond (\eg due to spoofing). Amplification increases drastically.}
  \label{fig:quic-ampl-factor-resend}
  \end{center}
\end{figure}

Our previous analysis considered the behavior of servers when handshakes complete successfully.
Now, we consider the case when a client fails to send an \texttt{ACK} to a server response.
This would cause a resend of data by the server.
Since the client IP~address is not verified all resends must comply in sum with the $3\times$ amplification limit.
A resend occurs, for example, when malicious actors initiate a handshake with a spoofed IP address.
All resends, \ie amplified traffic, are reflected to the spoofed address belonging to a victim.

\autoref{fig:quic-ampl-factor-resend} depicts amplification factors of handshakes collected at our telescope vantage point.
We sum all bytes received from a server for a specific SCID, and divide by an assumed client \texttt{Initial} of 1362 bytes.
All hypergiants exceed the amplification limit due to resends.
The majority of Cloudflare and Google backscatter traffic remains below factors of $10\times$.
Worryingly, traffic from Meta servers lead to amplification factors of up to $45\times$.
As a crosscheck, we inspect the duration of backscatter sessions for Meta.  We find a median of $\sim$51s and a maximum of 206s.
This indicates that the amplified traffic is received within a short time frame and the observed amplification factors are not biased by \eg reused, overlapping SCIDs.

To confirm that Meta servers do not comply with the current QUIC~specification~\cite{RFC-9000}, we conduct active scans as follows.
We send a single QUIC \texttt{Initial} but do not acknowledge the response.
We focus on the \texttt{/24} subnet of a Meta point-of-presence and identify three groups of IP addresses:
\begin{enumerate}
	\item No response or $\le$150 bytes, due to no QUIC HTTP3 service.
	\item Responses of $\approx$7k bytes, which corresponds to an amplification factor of over 5$\times$. IP addresses that typically serve \texttt{facebook.com} (\texttt{*.35, *.36}) belong to this group.
	\item Responses of $\approx$35k bytes, which corresponds to an amplification factor of over $28\times$.
	This amplification factor is similar to what can be achieved using popular amplification protocols~\cite{r-ahrnp-14}. We find IP~addresses that relate to Instagram and WhatsApp (\texttt{*.60, *.63}) belonging to this group.
\end{enumerate}

Overall, our active scans confirm the telescope observations.
Current deployments of Metas QUIC implementation \texttt{mvfst}~\cite{github-mvfst} do not respect the $3\times$ limit in case of resends.
Those deployments can be misused as amplifiers in attacks.

\section{Discussion \& Guidance}
\label{sec:discussion}

\paragraphNoDot{Should the QUIC protocol specification be updated?}
Our results suggest that the QUIC anti-amplification limit specified in RFC~9000~\cite{RFC-9000} is indeed tight but large enough to achieve 1-RTT handshakes.
The limit does not need to be increased to foster better deployments when network conditions are reliable.
In the case of packet loss and necessary resends, the anti-amplification limit challenges performance, though. 
It allows for at most one single retransmission of all server \texttt{Initial} and \texttt{Handshake} messages, given current certificate deployments including small ECDSA certificate chains and certificate compression.
Dealing efficiently with loss of messages during the connection setup seems an open challenge.

Next to protocol design challenges, we also find non-standard QUIC implementations that amplify during the 1-RTT handshake and increase significantly for incomplete QUIC handshakes.
More comprehensive testing of QUIC implementations is clearly~needed.

\paragraphNoDot{Does certificate compression help?}
We found that certificate compression is an impactful extension to allow servers staying below the amplification limit.
Unfortunately, popular TLS implementations such as OpenSSL do not support certificate~compression.
Given that recent QUIC implementations (\eg Microsoft QUIC) depend on existing TLS libraries, compression may remain in far reach and alternate measures are required to improve the situation.

\paragraphNoDot{Can a QUIC client mitigate lack of compression?}
To be independent of certificate compression, a QUIC~client could maintain a cache that includes certificate sizes of servers that the client frequently requests.
For entries in the cache, the client can then adapt the size of \texttt{Initial} requests to comply with the anti-amplification limit of the servers and achieve low latency connection setup.

\paragraph{Guidance for certificate authorities}
We argue that carefully created TLS certificates and certificate chains can positively influence the QUIC protocol performance.
ECDSA certificates lead to substantially smaller certificates chains. %
They can, however, not unfold their potential because especially root certificates are secured by RSA algorithms.
Our results show that updating these certificates can have beneficial cascading effects.

\paragraph{Guidance for QUIC implementations}
We infer the following guidelines when implementing QUIC network stacks:
First, at the server side implementation, bytes that result from padding or \texttt{Resend} must be included in anti-amplification limit checks.
Second, enabling packet coalescence at the server is recommended to omit padding and thus free space for TLS certificates reducing the need for additional round trips.
However, this can increase the latency when large-scale deployments deliver certificates by servers others than those providing content.
Third, we recommend the integration of a TLS library that supports compression to compensate large TLS~certificates, which currently trigger multi-RTT handshakes. %

\vspace{-0.5cm}

\section{Conclusion and Outlook}
\label{sec:conclusion}
   
In this paper, we measured and analyzed the QUIC handshake processes in the wild and found that the current Web certificate ecosystem challenges the QUIC design objective of a 1-RTT quick connection setup at low amplification potential.
As a consequence, large portions of QUIC connection setups are either multi-RTT, do not comply to the amplification limit, or both.
Future work shall closely monitor the evolution of the QUIC ecosystem and analyze the impact of measures to reduce certificate sizes effectively.

\paragraph{Responses from Hypergiants}
We contacted Meta as well as Cloudflare.
Details about our responsible disclosure policy are explained in \autoref{apx:ethical-concerns}.

\begin{acks}
We would like to thank our shepherd Olivier Bonaventure and the anonymous reviewers for their detailed and helpful feedback.
We gratefully acknowledge the wisdom of Gorry Fairhurst when discussing IETF~magic.
The assistance provided by Nicholas Banks during early stages of QUIC handshake classification is greatly appreciated.
This work was supported in parts by the \grantsponsor{BMBF}{German Federal Ministry of Education and Research (BMBF)}{https://www.bmbf.de/} within the projects Deutsches Internet-Institut (grant no. \grantnum{BMBF}{16DII111}) and \grantnum{BMBF}{PRIMEnet}.
\end{acks}

\label{lastpage}

\balance

\bibliographystyle{ACM-Reference-Format}
\bibliography{bibliography}

\begin{appendix}
  \section{Artifact Appendix}
\label{apx:artifacts}

\subsection{Abstract}
This section gives a brief overview of the artifacts of this paper.
We contribute tools to conduct follow-up  measurements as well as raw data and analysis scripts to reproduce the results and figures presented in this paper.

\subsection{Artifact check-list (meta-information)}

{\small
\begin{description}
  \item[Data set:] QUIC handshakes, TLS certificates. 
  \item[Run-time environment:] Python, Jupyter Notebooks.
  \item[Output:] All paper figures. 
  \item[How much disk space required?] 500MB
  \item[Time needed to prepare workflow?] 10 minutes
  \item[Time needed to complete experiments?] 20 minutes
  \item[Publicly available?] Yes
  \item[Archived?] Yes: \url{https://doi.org/10.5281/zenodo.7157904} 
\end{description}
}

\subsection{Description}

\subsubsection{How to access}

All artifacts are available via the following public repository:
\begin{itemize}[leftmargin=3mm]
  \item[]\url{https://github.com/ilabrg/artifacts-conext22-quic-tls}
\end{itemize}
This public repository provides up-to-date instructions for installing, configuring, and running our artifacts.
We also archive the camera-ready version of our software on Zenodo:
\begin{itemize}[leftmargin=3mm]
  \item[]\url{https://doi.org/10.5281/zenodo.7157904}
\end{itemize}

\subsubsection{Overview}
\begin{figure*}
	\input{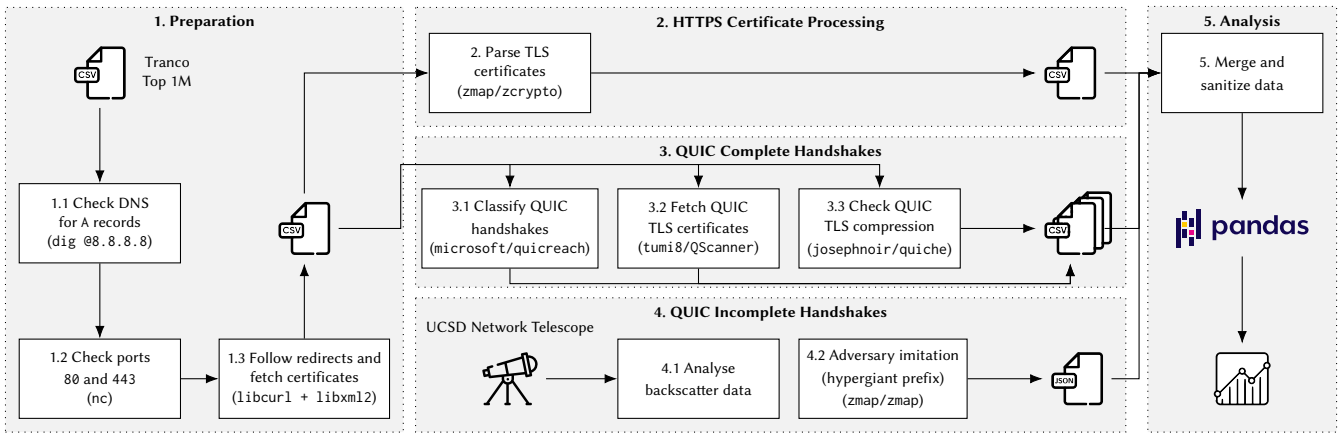}
	\caption{Overview of our HTTPS and QUIC analysis toolchain as well as related data flow.}
	\label{fig:toolchain}
\end{figure*}

We present an overview of our toolchain and related data flow in \autoref{fig:toolchain}.

\subsubsection{Software dependencies}
The minimal requirements to reproduce our figures are Python with Jupyter Notebook, Pandas, Matplotlib, Seaborn, and Zstd.
For a full list of dependencies, see the \texttt{requirements.md} file in the repository.

\subsubsection{Network dependencies}
We add all data to reproduce our results.
If you use the open-source tools to collect fresh data, your network has to allow fast connection setups on the default QUIC and HTTPS ports, \ie UDP/443 and TCP/443.

\subsubsection{Data sets}

We contribute two main data sets:
\one TLS certificate information of all 1M Tranco domains and related redirects and 
\two based on this list, QUIC handshakes to services with TLS certificates with varying \texttt{INITIAL} sizes.
Supporting data sets are documented in the repository.

\subsection{Installation}

Clone the artifacts repository, follow the \texttt{README} instructions to install the requirements, and then run Jupyter Notebook.
We provide different notebooks for different parts of the evaluation of the collected data.
No further installation is required.

\subsection{Evaluation and Expected Results}

We now explain the structure of the repository and which Python scripts should be run in which order.

\begin{description}
\item[\texttt{code/}]
Contains all analysis notebooks to reproduce our figures.
As a preparatory step, run \texttt{01-Prepare-Dataframes}, which parses and sanitizes the raw CSV data and prepares dataframes in compressed pickle format for fast reading.
Thereafter, you can execute any notebook in arbitrary order.

\item[\texttt{code/plots}]
Contains all figures in PDF and PNG format.
These files will be overwritten when rerunning the analyses.

\item[\texttt{data/csv}]
Raw, compressed CSV files from our active scans.
Large files have been split into multiple chunks to avoid running into file limits at \eg Github.

\item[\texttt{data/pkl/}]
Sanitized dataframes in compressed, binary format.
One file per dataframe.
These files will be created locally by running notebook \texttt{01-Prepare-Dataframes}.

\item[\texttt{misc/}]
Collection of open-source tools that we used to scan the Internet and to create the raw CSV files.
Some of these tools have been extended by us.
We contributed the extensions upstream but our pull request is not merged into the main branch of the third-party tools, yet. 
\end{description}

  \section{Ethical Concerns}
\label{apx:ethical-concerns}
This work may raise the following ethical concerns.

\paragraph{Educating attackers}
We discovered deployment behavior that conflicts with DDoS~mitigation required by RFC~9000 and hence enables misuse.
We follow a responsible disclosure policy and aim for fixing the bugs in collaboration with Cloudflare and Meta.

\begin{figure*}
	\begin{subfigure}{.99\textwidth}
		\begin{center}
		\includegraphics[width=\textwidth,keepaspectratio]{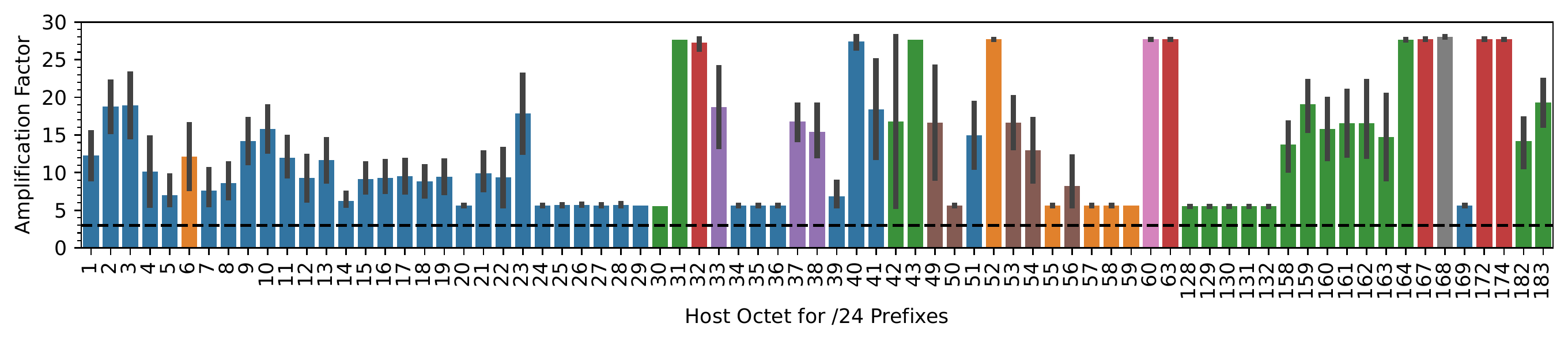}
		\caption{August 2022 (before disclosure)}
		\label{fig:facebook-amplification-fix-aug}
	    \end{center}
	\end{subfigure}
	\begin{subfigure}{.99\textwidth}
		\begin{center}
		\includegraphics[width=\textwidth,keepaspectratio]{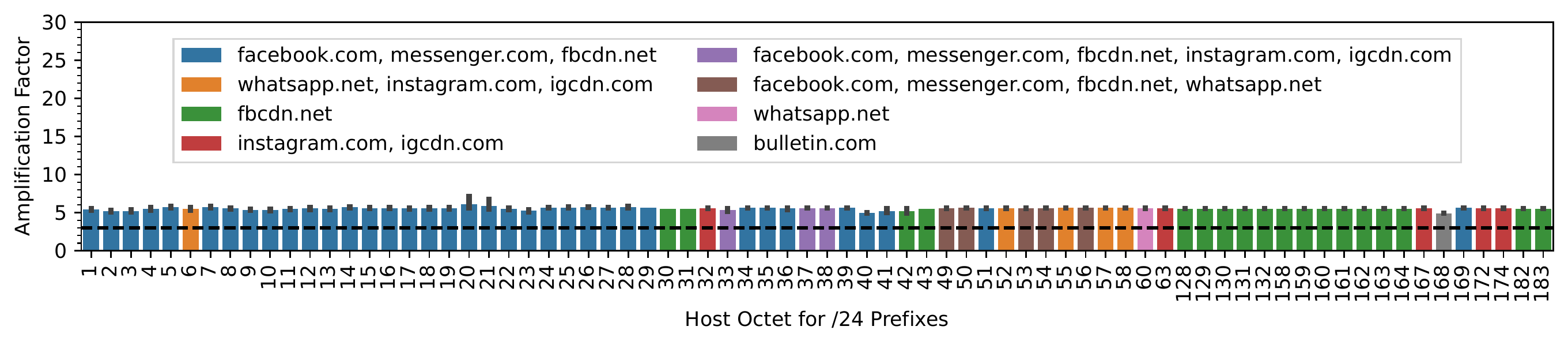}
		\caption{October 2022 (after disclosure)}
		\label{fig:facebook-amplification-fix-oct}		    
		\end{center}
	\end{subfigure}
	\caption{Mean amplification factors for Meta services observed at all point-of-presences. We see a significant improvement after the responsible disclosure of our results. The anti-amplification limit is still slightly above the allowed threshold.}
	\label{fig:facebook-amplification-fix}
\end{figure*}

\paragraph{Responses from hypergiants}
\label{apx:reaction-hypergiants}
Meta significantly improved their QUIC deployment in October 2022.
By rescanning all host addresses in \texttt{/24} on-net prefixes, we now observe homogeneously configured servers that limit the amount of QUIC retransmissions in case of unverified clients.
However, with a mean amplification factor of 5$\times$, the responses still exceed the anti-amplification limit specified in RFC~9000.
We show the results in \autoref{fig:facebook-amplification-fix}, including 95\% confidence intervals.

Cloudflare has responded, and explains the reason to exceed the limit is to help improve client performance, while respecting production constraints that are omitted from the QUIC specification.
Specifically, in production environments the information needed to populate the \texttt{ServerHello} is contained in certificates that may be managed separately from connection termination, and unavailable at the moment of arrival of the client’s \texttt{Initial}.
The delay affects client estimates of RTT.
Cloudflare mitigates the delay by immediately responding to client \texttt{Initials} with an \texttt{ACK} padded at the UDP layer.
This occurs once, so the amplification factor is bound.

\section{QUIC Anti-Amplification Limit}
\label{apx:ietf-limit}

In \autoref{tbl:ietf-ampl-limit}, we show the historical development of QUIC amplification mitigation as proposed in the different versions of the QUIC~Internet~Draft.
Although amplification attacks have already been mentioned in Draft~01~\cite{draft-ietf-quic-transport-01}, no limitations to reduce the attack potential have been specified for servers.
Draft~02~\cite{draft-ietf-quic-transport-02}, at least, specifies that clients must ensure that the first packet in a connection, \ie commonly an \texttt{INITIAL}, meets the requirement of minimum packet size.
This requirement limits the overall amplification factor since any attacker needs to invest a minimum amount of data.

In Draft~09~\cite{draft-ietf-quic-transport-09}, the first restriction for servers is introduced.
A server may close a connection with an error code in the case of a too small client \texttt{INITIAL}.
Otherwise, it must not respond or behave as if any part of the offending packet was processed as valid.
In Draft~10~\cite{draft-ietf-quic-transport-10}, a server is limited by the number of \texttt{HANDSHAKE} packets a server is allowed to send to unverified clients, even though this is not explicitly noted in the context of amplification mitigation.
Since Draft~15~\cite{draft-ietf-quic-transport-15}, the anti-amplification limit is specified relative to the client.
Since Draft~33~\cite{draft-ietf-quic-transport-33}, including the current RFC~\cite{RFC-9000}, this limit has been specified to three times of received \textit{data}.

We find little discussion about the limit on the IETF mailing lists.
In March 2018, 3600 ($= 3\cdot1200$) bytes have been discussed as ``decently large''~\cite{mails-ietf-3600} to carry TLS certificates.
A recent question on the exact motivation behind the 3$\times$ remains unanswered~\cite{mails-ietf-3x}.

\begin{table}[t]
	\caption{Descriptions of amplification mitgation in the different versions of the IETF QUIC Internet Draft, leading to the 3$\times$ anti-amplification limit. [Bold highlighting by us.]}
	\label{tbl:ietf-ampl-limit}
	\footnotesize
	\begin{tabularx}{\columnwidth}{p{1.27cm} r X}
		\toprule
        IETF Spec & \makecell[c]{Date} & Proposed Limit \\
		\midrule
	    Draft 09 & 01/2018 & ``A server MAY send a CONNECTION\_CLOSE frame with error code PROTOCOL\_VIOLATION in response to an Initial packet smaller than 1200 octets.'' \\
	    \dcline{1-3}
		Draft 10 -- 12 & 03/2018 -- 05/2018 & ``Servers MUST NOT send more than three \textbf{Handshake} packets without receiving a packet from a verified source address.'' \\
	    \dcline{1-3}
	    Draft 13 -- 14 & 06/2018 -- 08/2018 & ``Servers MUST NOT send more than three \textbf{datagrams} including Initial and Handshake packets without receiving a packet from a verified source address.'' \\
	    \dcline{1-3}
	    Draft 15 -- 32 & 10/2018 -- 10/2020 & ``Servers MUST NOT send more than three times as many \textbf{bytes} as the number of bytes received prior to verifying the client's address.'' \\
	    \dcline{1-3}
	    \makecell[lt]{Draft 33 -- 34, \\RFC 9000} & \makecell[rt]{12/2020 -- 01/2021, \\05/2021} & ``[\ldots] an endpoint MUST limit the amount of \textbf{data} it sends to the unvalidated address to three times the amount of data received from that address.'' \\
		\bottomrule
	\end{tabularx}
\end{table}

\section{Influence of Top List Ranks}
\label{apx:influence-rank}

We verify whether our results depend on some type of popularity of the QUIC-based Web service using the Tranco list~\cite{LePochat2019tranco}.
To this end, we split the Tranco list in groups of 100k (ranked) names and initiate QUIC and HTTPS handshakes for each name.

\begin{figure}
  \begin{center}
  \includegraphics[width=1\columnwidth]{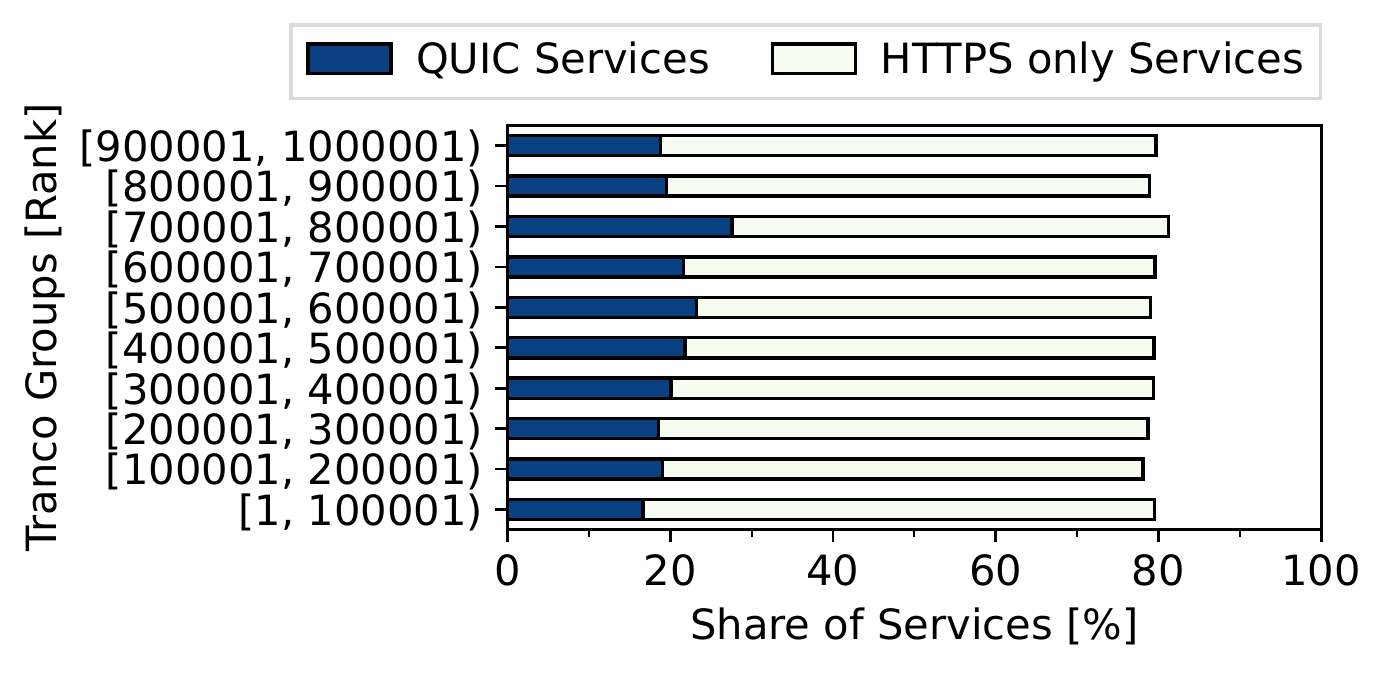}
      \caption{Service popularity across tranco rank groups. QUIC and HTTPS deployment rates are stable across rank groups.}
  \label{fig:quic-reachable-sites-rank}
  \end{center}
\end{figure}

\autoref{fig:quic-reachable-sites-rank} exhibits the relative amount of servers that are reachable via QUIC or only via HTTPS.
On average, 21\% of domains per rank group are reachable via QUIC.
On top of this, $\approx59\%$ of additional names own a TLS certificate and are reachable over HTTPS.
The popularity of a server has no influence on the popularity of QUIC deployment, as we observe a small standard deviation of $\sigma=3$ across rank groups.

\begin{figure}%
  \begin{center}
  \includegraphics[width=1\columnwidth]{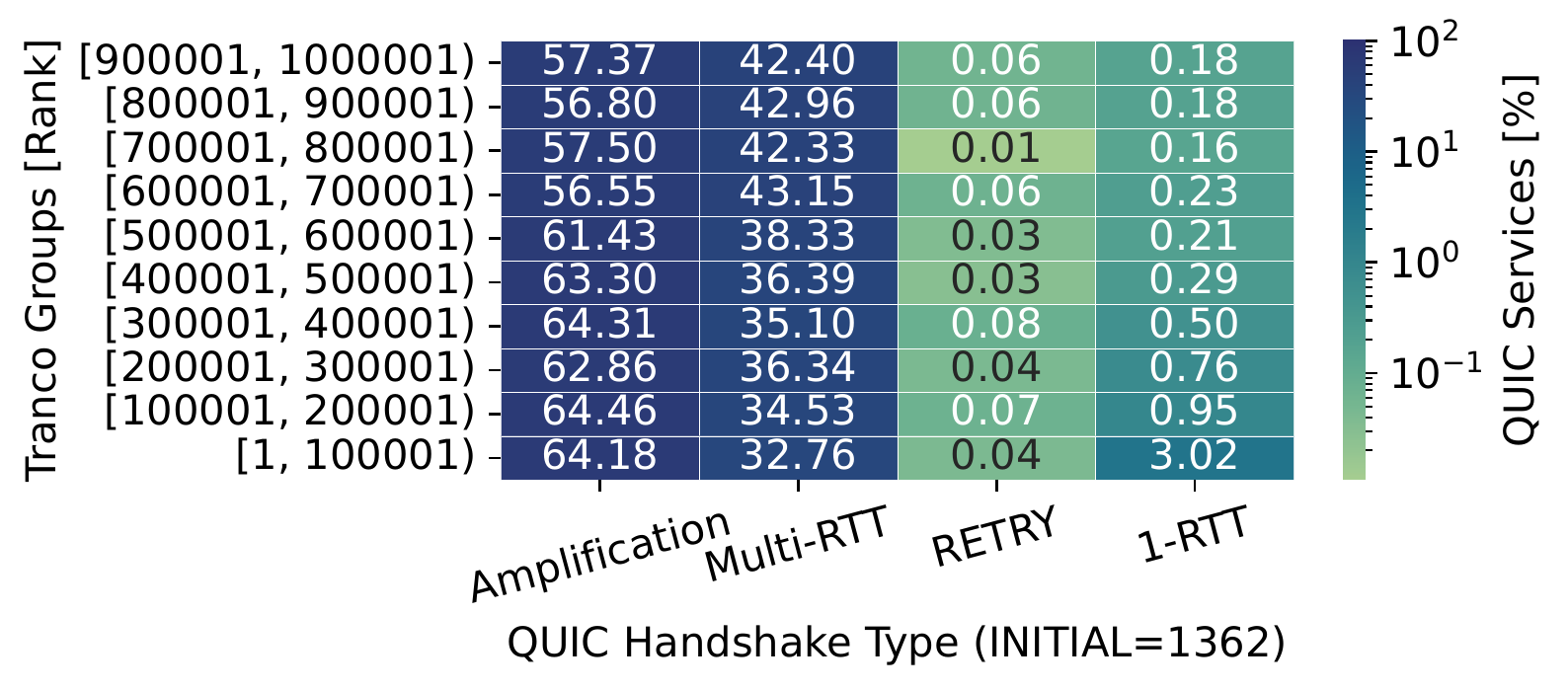}
      \caption{QUIC handshake classification per tranco rank group. Handshake types are mostly stable across rank groups.}
  \label{fig:quic-handshake-classes-rank}
  \end{center}
\end{figure}

We also check whether the QUIC handshake classification is stable across ranks by mapping responses to a QUIC handshake type (amplification, multi-RTT \etc) and counting the relative number of servers per type.
\autoref{fig:quic-handshake-classes-rank} visualizes the results.
Again, we find no significant differences across rank groups.
The only exceptions are 1-RTT handshakes, which appear more popular among the 100k most popular QUIC servers (3.02\% vs. <0.95\%).

Both analysis indicate that our results are independent of the specific Tranco rank.

\section{Cruise-Liner TLS Certificates}
\label{apx:cruise-liner}

\begin{figure}[b]
  \begin{center}
  \includegraphics[width=1\columnwidth]{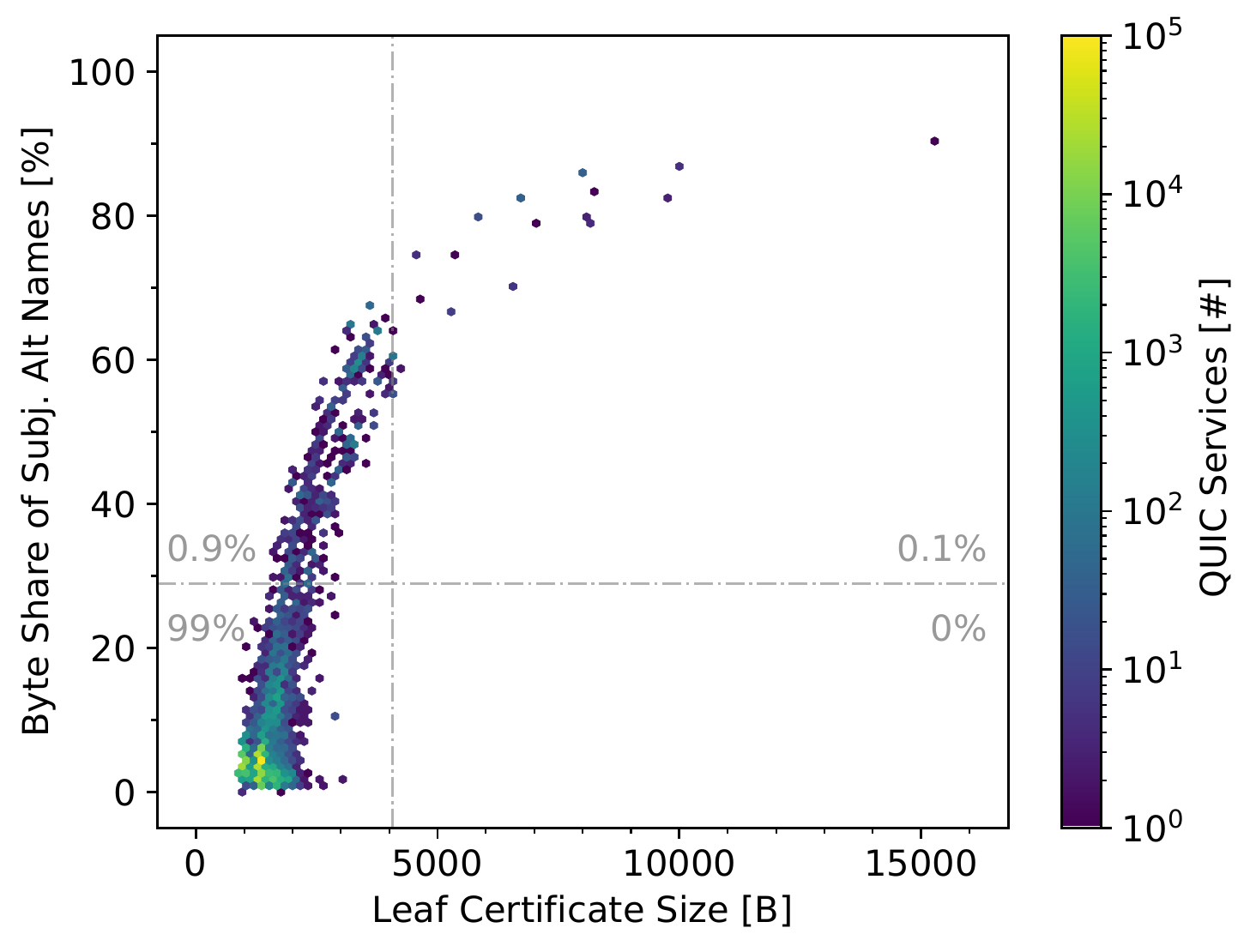}
      \caption{Relative size of subject alternative names (SANs). Cruise-liner certificates are rare for QUIC services.}
  \label{fig:cruise-liner-certs}
  \end{center}
\end{figure}

Cruise-liner certificates~\cite{cangialosi2016httpskeys} are certificates that are large in size due to many subject alternate names (SANs).
We now check whether QUIC services are affected by cruise-liner certificates.
To this end, we analyze all the leaf certificates received for all QUIC services.
We inspect the total certificate size and the share of bytes required by all SANs.
The results are visualized in \autoref{fig:cruise-liner-certs}.

Overall, most SANs amount for less than 10\% of bytes.
Taking a closer look at the top 1\% of certificates by SAN byte share, we find that they require at least 28.9\% of bytes (horizontal threshold).
Worryingly, 0.1\% of certificates exhibit a high SAN byte share and exceed a common QUIC amplification limit (vertical threshold).

\end{appendix}

\end{document}